\documentclass{llncs}

\newcommand{\Oh}[1]{\ensuremath{\mathcal{O}\!\left({#1}\right)}}
\newcommand{\parse}{\ensuremath{\mathsf{parse}}}

\begin{document} 

\title{Approximating LZ77 in Small Space}
\author{Travis Gagie}
\institute{Department of Computer Science, University of Helsinki, Finland}	
\maketitle

\section{Introduction}
\label{sec:introduction}

Although nearly forty years old, Ziv and Lempel's LZ77~\cite{ZL77} is still one of the best algorithms for compressing repetitive datasets such as genomic databases, software repositories and versioned text.  The LZ77 parse of a string can also be used to speed up online pattern-matching and to reduce the size of indexes; see, e.g.,~\cite{GP15}.  When the dataset is too large to fit in internal memory, however, computing the parse is a challenge.  The best known external-memory algorithm~\cite{KKP14} for computing the parse exactly takes $\Oh{n^2 / (M B)}$ I/Os, where $M$ is the size of the internal memory and $B$ is the size of a disk block.  There are more efficient external-memory algorithms that can be used to approximate the parse --- see, e.g.,~\cite{MT14} --- but the known approximation ratios are at least logarithmic.

In this paper we describe a randomized algorithm with which, given a positive \(\epsilon \leq 1\) and read-only access to a string \(S [1..n]\) whose LZ77 parse consists of $z$ phrases, with high probability we can build an LZ77-like parse of $S$ that consists of $\Oh{z / \epsilon}$ phrases using $\Oh{n^{1 + \epsilon}}$ time, $\Oh{n^{1 + \epsilon} / B}$ I/Os and $\Oh{z / \epsilon}$ space.  By ``LZ77-like'' we mean that, first, the concatenation of the phrases is $S$; second, for each phase \(S [i..j]\), either \(i = j\), in which case we encode \(S [i]\) as \(\langle 0, S [i] \rangle\), or \(S [i..j]\) first occurs at position $i'$ in \(S [1..j - 1]\), in which case we encode \(S [i..j]\) as \(\langle i', j - i + 1 \rangle\).

\section{Algorithm}
\label{sec:algorithm}

Assume we have an easy way to determine where substrings of lengths \(n, f (n),\) \(f (f (n)), \ldots, 2\) first occur in $S$, where \(f (\ell) = \min (\lceil \ell (1 - 1 / n^\epsilon) \rceil, \ell - 1)\).  Then we can approximate the LZ77 parse by calling \(\parse (1, n, n)\), where \(\parse (i, j, \ell)\) is the following recursive procedure:
\begin{enumerate}
\item if \(\ell = 1\) then we encode \(S [i], \ldots, S [j]\) as \(\langle 0, S [i] \rangle, \ldots, \langle 0, S [j] \rangle\) and return;
\item if \(j - i + 1 < \ell\) then we call \(\parse (i, j, f (\ell))\) and return;
\item if \(S [j - \ell + 1..j]\) first occurs at position $i'$ in \(S[1..j - 1]\), then we call \(\parse (i, j - \ell, \ell)\), encode \(S [j - \ell + 1..j]\) as \(\langle i', \ell \rangle\), and return;
\item for $k$ from 0 to \(\lfloor (j - i + 1) / \ell \rfloor - 1\), if \(S [i + k \ell..i + (k + 1) \ell - 1]\) first occurs at position $i'$ in \(S [1..i + (k + 1) \ell - 2]\), then we call \(\parse (i, i + k \ell - 1, f (\ell))\), encode \(S [i + k \ell..i + (k + 1) \ell - 1]\) as \(\langle i', \ell \rangle\), call \(\parse (i + (k + 1) \ell, j, \ell)\), and return;
\item if none of the conditions above are met, then we call \(\parse (i, j, f (\ell))\) and return.
\end{enumerate}

For example, suppose \(S = \mathtt{ababbabbaabbabbaababa}\) and \(21^\epsilon = 4\), so \(f (\ell) = \min (\lceil 3 \ell / 4 \rceil, \ell - 1)\):
\begin{enumerate}
\item \(\parse (1, 21, 21)\) calls \(\parse (1, 21, 16)\);
\item \(\parse (1, 21, 17)\) calls \(\parse (1, 21, 12)\);
\item \(\parse (1, 21, 14)\) calls \(\parse (1, 21, 9)\);
\item \(\parse (1, 21, 9)\) calls \(\parse (1, 9, 9)\),
\newline encodes \(S [10..18]\) as \(\langle 3, 9 \rangle\),
\newline and calls \(\parse (19, 21, 9)\);
\item \(\parse (1, 9, 9)\) calls \(\parse (1, 9, 7)\);
\item \(\parse (1, 9, 7)\) calls \(\parse (1, 9, 6)\);
\item \(\parse (1, 9, 6)\) calls \(\parse (1, 9, 5)\);
\item \(\parse (1, 9, 5)\) calls \(\parse (1, 4, 5)\),
\newline and encodes \(S [5..9]\) as \(\langle 2, 5 \rangle\);
\item \(\parse (1, 4, 5)\) calls \(\parse (1, 4, 4)\);
\item \(\parse (1, 4, 4)\) calls \(\parse (1, 4, 3)\);
\item \(\parse (1, 4, 3)\) calls \(\parse (1, 4, 2)\);
\item \(\parse (1, 4, 2)\) calls \(\parse (1, 2, 2)\)
\newline and encodes \(S [3..4]\) as \(\langle 1, 2 \rangle\);
\item \(\parse (1, 2, 2)\) calls \(\parse (1, 2, 1)\);
\item \(\parse (1, 2, 1)\) encodes \(S [1]\) as \(\langle 0, \mathtt{a} \rangle\),
\newline and encodes \(S [2]\) as \(\langle 0, \mathtt{b} \rangle\);
\item \(\parse (19, 21, 9)\) calls \(\parse (19, 21, 7)\);
\item \(\parse (19, 21, 7)\) calls \(\parse (19, 21, 6)\);
\item \(\parse (19, 21, 6)\) calls \(\parse (19, 21, 5)\);
\item \(\parse (19, 21, 5)\) calls \(\parse (19, 21, 4)\);
\item \(\parse (19, 21, 4)\) calls \(\parse (19, 21, 3)\);
\item \(\parse (19, 21, 3)\) encodes \(S [19..21]\) as \(\langle 1, 3 \rangle\).
\end{enumerate}
Thus, our parse is
\[\mathtt{a}, \mathtt{b}, \mathtt{ab}, \mathtt{babba}, \mathtt{abbabbaab}, \mathtt{aba}\,,\] encoded as
\[\langle 0, \mathtt{a} \rangle, \langle 0, \mathtt{b} \rangle, \langle 1, 2 \rangle, \langle 2, 5 \rangle,  \langle 3, 9 \rangle, \langle 1, 3 \rangle\,,\]
while the true LZ77 parse is
\[\mathtt{a}, \mathtt{b}, \mathtt{abb}, \mathtt{abbaa}, \mathtt{bbabbaaba}, \mathtt{ba}\,.\]

It is not difficult to verify that our algorithm terminates and, assuming we find the first occurrences of substrings correctly, produces a valid encoding.  The interesting problems are, first, to show how we can easily determine where substrings first occur in $S$ and, second, to analyse its performance.  Notice that if we make a recursive call on an interval \(S [i..j]\), the only substrings of length $\ell$ we ever consider in that interval are in
\begin{eqnarray*}
&& \{ S [i + k \ell..i + (k + 1) \ell - 1]\ :\ 0 \leq k \leq \lfloor (j - i + 1) / \ell \rfloor - 1 \}\ \cup \\
&& \{ S [j - (k + 1) \ell + 1..j - k \ell]\ :\ 0 \leq k \leq \lfloor (j - i + 1) / \ell \rfloor - 1 \}\,.
\end{eqnarray*}

If we compute the Karp-Rabin fingerprints~\cite{KR87} of those substrings and store them in a membership data structure, such as a hash table, then we can search for their first occurrences by passing a sliding window of length $\ell$ over $S$, maintaining the fingerprint of the window and performing a membership query for it at each step.  In fact, if we keep the branches of the recursion synchronized properly, we can use a single membership data structure and a single pass of a sliding window to find the first occurrences of all the substrings of length $\ell$ we ever consider while parsing $S$.  Of course, using Karp-Rabin fingerprints means our algorithm is randomized and could produce an invalid encoding, albeit with a very low probability.  We can check an encoding by comparing each phrase to the earlier substring it should match.

In our example, step 1 requires passes with sliding windows of length 21 and 16 before we can execute \(\parse (1, 21, 21)\) and \(\parse (1, 21, 16)\); step 2, with one of length 12; step 3, with one of length 9; step 4 does not require a pass; steps 5 and 15, with one of length 7; steps 6 and 16, with one of length 6; steps 7 and 17, with one of length 5; step 8 does not require a pass; steps 9 and 18, with one of length 4; steps 10 and 19, with one of length 3; step 11, with one of length 2; steps 12, 13, 14 and 20 do not require a pass.

A practical optimization would be to build a table that stores the position of the first occurrence of each substring of $S$ of length \(f^{[i]} (n) < \log_\sigma M - c\), where $M$ is the size of the internal memory and $c$ is a constant.  Building this table takes $\Oh{n \log_\sigma M}$ time, one pass over $S$ and a \(1 / \sigma^c\) fraction of the internal memory.  It allows us to avoid making passes with sliding windows shorter than \(\log_\sigma M - c\).

\section{Time, I/O and Space Bounds}
\label{sec:bounds}

Our algorithm's time and I/O complexities are dominated by the time and I/Os needed to make a pass over $S$ for each length \(n, f (n), f (f (n)), \ldots, 2\).  If we use a perfect hash table for the fingerprints, the time for each pass is $\Oh{n}$, and calculation shows that the number of passes is \(\Oh{\log_{\frac{1}{1 - 1 / n^\epsilon}} n} = \Oh{n^\epsilon \log n}\).  Therefore, we use $\Oh{n^{1 + \epsilon} \log n}$ total time and $\Oh{n^{1 + \epsilon} \log (n) / B}$ total I/Os.

The space complexity is dominated by the data structures for handling the recursion itself, and by the hash tables.  The total size of the data structures for the recursion can be bounded in terms of the number of phrases in our parse, which we show in the next section to be $\Oh{z / \epsilon}$.  We need only store one hash table at a time, and its size is proportional to the number of substrings we are currently considering.  We consider a substring only if the all larger substrings overlapping it do not occur earlier in $S$, meaning they cross or end at a bounary between phrases in the LZ77 parse.  It follows that each substring of length $\ell$ we consider must be within distance $\frac{\ell}{1 - 1 / n^\epsilon}$ of a phrase boundary.  Therefore, we consider $\Oh{z}$ substrings of any given length.

\section{Approximation Ratio}
\label{sec:ratio}

Consider how many phrases in our parse can overlap a phrase \(S [i..j]\) in the LZ77 parse.  The longest phrase we use that overlaps \(S [i..j]\) either completely covers \(S [i..j]\), covers a proper prefix of it, covers a proper suffix of it, or is properly contained in it.  The first case is trivial; the second and third cases are essentially symmetric, as we are left trying to bound the number of our phrases that can overlap either a proper suffix or a proper prefix of \(S [i..j]\) that is immediately preceded or followed by one of our phrases; and the fourth case is a combination of the second and third, as we are left trying to bound the number of our phrases that can overlap a proper prefix and a proper suffix of \(S [i..j]\) that are separate by one of our phrases.  

Without loss of generality, we consider only the second case.  That is, assume \(S [i..i' - 1]\) is covered by one of our phrases, for some \(i' > i\), and consider how many of our phrases can overlap \(S [i'..j]\).  By inspection of our algorithm, the longest phrase we use that overlaps \(S [i'..j]\) either covers a suffix of it, say \(S [j' + 1..j]\), or starts at $i'$ and covers at least a \(1 - 1 / n^\epsilon\) fraction of \(S [i'..j]\).

In the first case now, the longest phrase we use that overlaps \(S [i'..j']\) must end at \(S [j']\) and cover at least a \(1 - 1 / n^\epsilon\) fraction of \(S [i'..j']\).  The second case is essentially the same as what we just considered.  It follows that the number of our phrases that can overlap an LZ77 phrase is \(\Oh{\log_{n^\epsilon} n} = \Oh{1 / \epsilon}\), so we use $\Oh{z / \epsilon}$ phrases in total.

We can summarize the previous section and this one so far as follows: given \(\epsilon > 0\) and read-only access to a string \(S [1..n]\) whose LZ77 parse consists of $z$ phrases, with high probability we can build an LZ77-like parse consisting of $\Oh{z / \epsilon}$ phrases using $\Oh{n^{1 + \epsilon} \log n}$ time, $\Oh{n^{1 + \epsilon} \log (n) / B}$ I/Os and $\Oh{z / \epsilon}$ space.  Notice that, if we divide $\epsilon$ by 2 before running our algorithm, then the time and I/O bounds are \(\Oh{n^{1 + \epsilon / 2} \log n} = \Oh{n^{1 + \epsilon}}\) and \(\Oh{n^{1 + \epsilon / 2} \log (n) / B} = \Oh{n^{1 + \epsilon} / B}\), so we get the slightly neater results stated in the introduction. 

\begin{theorem}
\label{thm:main}
Given a positive \(\epsilon \leq 1\) and read-only access to a string \(S [1..n]\) whose LZ77 parse consists of $z$ phrases, with high probability we can build an LZ77-like parse of $S$ that consists of $\Oh{z / \epsilon}$ phrases using $\Oh{n^{1 + \epsilon}}$ time,
\linebreak $\Oh{n^{1 + \epsilon} / B}$ I/Os and $\Oh{z / \epsilon}$ space.
\end{theorem}

\end{document}